# Atomic-resolution elemental mapping at cryogenic temperatures enabled by direct electron detection

Berit H. Goodge[1], David J. Baek[2], and Lena F. Kourkoutis[1,3]

1. School of Applied and Engineering Physics, Cornell University, Ithaca, NY 14853, USA
2. School of Electrical and Computer Engineering, Cornell University, Ithaca, NY 14853, USA
3. Kavli Institute at Cornell for Nanoscale Science, Cornell University, Ithaca, NY 14853, USA

**Spectroscopic mapping by scanning transmission electron microscopy coupled with electron energy loss spectroscopy (STEM-EELS) is a powerful technique for determining the structure and chemistry of a wide range of materials and interfaces. The extension of this technique to cryogenic temperatures opens the door to new experiments across many fields including materials physics, energy storage and conversion, and biology. Such experiments, however, often face signal limitations due to sample sensitivity or the need for rapid data acquisition under less stable cryogenic conditions. Compared to traditional indirect detection systems such as charge coupled devices (CCDs), direct electron detectors (DEDs) offer improved detective quantum efficiencies, narrower point spread functions, and superior signal-to-noise ratios. Here, we compare the performance of a Gatan K2 Summit DED to an UltraScan 1000 CCD for use in signal-limited atomic-resolution STEM-EELS experiments. Due to its improved point spread function, the DED's energy resolution remains comparable to that of the CCD at a 5 times lower dispersion, providing simultaneous access to a much broader total energy range without sacrificing spectral resolution. More importantly, the benefits of direct detection enable a variety of low-signal experiments, including atomic-resolution mapping of minor and high energy edges such as the La-$M_{2,3}$ edge at 1123 eV and the Bi-$M_{4,5}$ edge at 2580 eV. For rapid acquisitions at 400 spectra per second, elemental maps recorded with the DED show an up to 40% increase in atomic lattice fringe contrast compared to those acquired with the CCD. Taking advantage of these performance improvements and the fast readout of the K2 DED, we use direct detection STEM-EELS to demonstrate atomic-resolution elemental mapping at cryogenic temperatures.**

# INTRODUCTION

Electron energy loss spectroscopy (EELS) performed with a tightly focused electron beam which is scanned across the sample in a scanning transmission electron microscope (STEM) has enabled spectroscopic mapping of a wide range of materials and interfaces[1–5]. Directly measuring interactions between primary electrons in the STEM probe and the bound electrons in the sample provides access to rich chemical information down to the atomic scale[6–8]. Compared to other bulk spectroscopic techniques, STEM-EELS can regularly achieve near-Å resolution, enabling, for example, detailed quantification of compositional gradients[9], dopant distribution[10], and of interfaces in atomically engineered heterostructures[11]. Additionally, the detailed fine-structure of an EEL spectrum contains rich chemical information which can be used to map subtle variations such as local bonding[12–15] and valence states[16,17].

In most contemporary experiments, a sub-Å STEM probe is rastered across an electron-transparent sample. Electrons from the beam that interact inelastically with those in the sample (for instance, by losing energy to excite core electrons into an excited state) are then dispersed in energy in a set of post-specimen electromagnetic lenses. The resulting spectrum is collected at the end of the spectrometer, historically by film plates but now most commonly by an electronic detector. The quantitative analytic capabilities of STEM-EELS have benefitted in recent decades from a number of both hardware and software developments, including aberration-corrected STEM lenses[8,18], improved spectrometers[19], and data processing[20,21]. In this report, we discuss recent instrumentation improvements to the EELS detector[22,23] and the new possibilities for materials research now within reach. In particular, these detector improvements result in enhanced contrast between spectral signal and background, which is especially important for weak or low-signal experiments. "Low-signal" experiments are those in which the number of electrons that contribute to a given EELS edge is significantly limited by any of a wide range of experimental factors, including low beam currents used for dose-sensitive samples, weakly scattering low cross-section edges, or extremely short acquisition times. Overcoming these limitations is critical for expanding high-resolution EELS to a wide range of systems, including biologic specimens, 2D materials, and variable-temperature experiments with reduced mechanical stability.

The most commonly used EELS detectors use parallel-recording systems, typically a scintillating material optically coupled to a charge-coupled device (CCD) photodetector. These systems are referred to as indirect detectors because the inelastically scattered electrons are first converted to photons in the scintillator which are transmitted through fiber optics to the CCD. These photons in turn excite electron-hole pairs in the diode array that create the electronic signal ultimately read out by the detector. Each stage of the detection process, however, degrades the quality of the recorded spectrum, measured both by point spread function (PSF) and signal to noise ratio (SNR)[24,25]. The PSF of a conventional detector can suffer broadening in two major stages: first because of lateral scattering in the scintillator material, and second from "blooming" (charge spillover to neighbouring pixels) in the CCD array. Similarly, each stage of detection also introduces a new source of noise, first as gain noise in the electron to photon conversion and then again during the CCD readout process[22,25].

In contrast, the DED consists of a single thin active epilayer in which electrons are sensed and counted directly. Eliminating both stages of electron-photon conversion helps to remove lateral scattering effects and conversion noise in the DED[22,26,27], resulting in improved detector quantum efficiency (DQE), narrower PSF, and superior SNR[22,23,26,27]. Here, we demonstrate the use of a Gatan K2 Summit direct electron detector for atomic-resolution spectroscopic mapping under a variety of signal-limited conditions and compare its performance to a Gatan UltraScan1000 traditional indirect detection CCD. The superior detector characteristics of the DED as well as a 4-fold reduction in read out time allow us to achieve atomic-resolution elemental mapping at cryogenic temperature which has previously been hampered by stage drift under cryogenic conditions.

**MATERIALS AND METHODS**

Both the K2 Summit and UltraScan 1000 cameras are installed at the end of a Gatan 965 GIF Quantum ER on an aberration-corrected FEI Titan Themis 60-300. The K2 DED inserts in front of the UltraScan CCD, allowing for easy switching between detectors. Experimental comparisons between the two detectors were performed by repeating identical acquisitions on either camera immediately following one another to minimize any changes in probe or sample conditions. All spectrum images (SIs) and core-loss data in this report were acquired with a sub-Ångström probe. The only correction applied to any data was high quality dark reference subtraction performed after each acquisition; no other drift correction, filtering, or binning has been used.

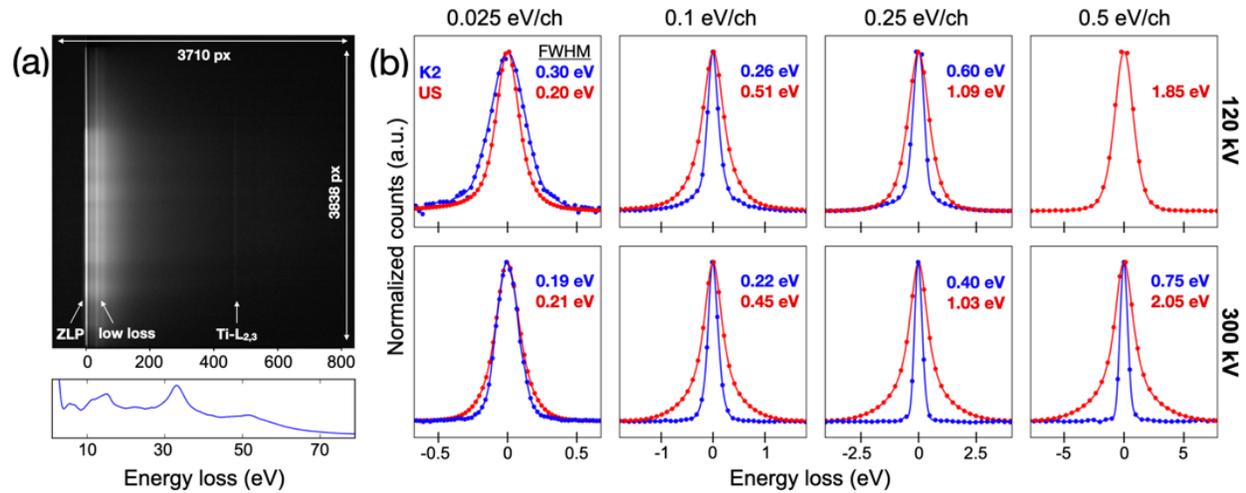

**Figure 1** | Spectra are recorded on the K2 DED by averaging the counts along 3838 pixels in each of the 3710 energy loss channels. (a) Camera image of the energy loss spectrum taken with 0.25 eV/ch dispersion over SrTiO$_3$ is dominated by the bright ZLP line, with low loss features and the Ti-L$_{2,3}$ edge visible at higher energies. (bottom) High SNR spectrum of the low loss region recorded by moving the ZLP off the detector. (b) Normalized ZLPs of a monochromated 120 kV (top) and 300 kV (bottom) TEM beam over vacuum with 0.025, 0.1, 0.25, and 0.5 eV/ch measured as the sum of 4000 auto-aligned frames on the UltraScan CCD (red) and K2 DED (blue). Spline curve fits are added to the raw data (dots) as guides to the eye. The full width at half maximum (FWHM) is given for each spectrum.

Beyond the detection process, the K2 Summit also differs from the UltraScan purely in size, increasing to 3710×3838 px from 2048×2048 px, respectively. Operated in spectroscopy mode, the K2 averages signal along the 3838 px direction to give a total of 3710 energy channels as shown in Figure 1a, an 81% increase over 2048 channels on the UltraScan. The bright zero loss peak (ZLP) line dominates a K2 camera image taken with 0.25 eV/ch dispersion over SrTiO$_3$ (STO), with low-loss band transition edges (top image and bottom axis) and the Ti-L$_{2,3}$ core-loss edge visible at higher energies (top image). Originally used for very low electron doe TEM imaging[28], the collection efficiency of the K2 begins to suffer from coincidence losses near 20 electrons per pixel per second[29]. When the energy loss spectrum is spread over the full vertical range (3838 px) of the detector, however, the coincidence loss threshold is equivalent to ~192 counts per channel, well above the levels for normal core-loss spectroscopy using typical beam currents (up to hundreds of pA). Depending on the size of spectrometer entrance aperture used, the maximum counts per channel must be scaled by the fraction of the detector being illuminated such that smaller entrance apertures will require a corresponding reduction in probe current to prevent saturation. Saturation from very strong signals such as the ZLP and the low-loss region can also be avoided by reducing the duty cycle with an electrostatic shutter to as low as 0.041[22,26,27].

**RESULTS**

By eliminating both stages of electron-photon conversion and instead detecting electrons directly in one thin active layer, the DED achieves a narrower PSF than the CCD. Operated in TEM mode, a 300 kV beam is monochromated to a nominal energy spread of 150 meV as measured on the CCD at 0.01 eV/ch dispersion. Spectra of the ZLP were collected by summing 4000 frames over vacuum with 0.025, 0.1, 0.25, and 0.5 eV/ch dispersions on both the CCD and DED.

At the highest dispersion (0.025 eV/ch), energy resolution is limited mostly by spectrometer tuning and instability rather than by the performance of either detector, both of which recorder very similar ZLPs. At lower dispersions (0.25 and 0.5 eV/ch), detector PSF plays a larger role and the Gaussian full width at half maximum (FWHM) is decreased by as much as half for the same conditions on the DED compared to the CCD. PSF improvements in the DED are more significant at the higher accelerating voltage of 300 kV (bottom row) compared to 120 kV (top row).

The difference in signal spreading on the two detectors is also notable in the different characteristic shape of each ZLP. When the beam energy spread is much greater than the energy per pixel (e.g.: 0.025 eV/ch dispersion for 150 meV beam spread), both detectors record very similar ZLPs. At lower dispersions, however, the DED data takes on a quasi-Gaussian form, while the CCD data is better fit by Lorentzians. Qualitatively, this translates to noticeably broader ZLP tails on the CCD than on the DED. We estimate the localization of signal on each detector in pixels by calculating the FWHM of the ZLP and dividing by energy per channel. On the DED, signals are localized within about 3-4 pixels, while on the CCD this quantity spreads over 5-6 pixels. Together with the different pixel densities of the two cameras, an average PSF delocalization of 3.5 pixels out of 3710 channels on the DED is an approximately 2.5 times improvement compared to a delocalization of 5.5 pixels out of 2048 channels on the CCD.

The effects of this improved PSF are particularly apparent when comparing energy resolution across a large simultaneous energy range, where closely-separated peaks risk blurring into one another. Figure 2 shows spectra collected in STEM mode without monochromation (operating energy resolution of ~1 eV) on the two detectors, demonstrating the advantage of the DED in terms of both energy resolution as well as accessible range. Operating the DED at the lowest dispersion of 0.5 eV per channel retains energy resolution of the core-loss edges comparable to that of the CCD at a much higher dispersion of 0.1 eV/ch, as evidenced most clearly by the four distinct peaks of the Ti-$L_{2,3}$ edge seen in the left of Figure 2. Narrow PSF, reduced noise, and high number of pixels allow the DED to simultaneously capture fine structure information from the Ti-$L_{2,3}$ edge at 455 eV while still distinctly resolving the Sr-$L_{2,3}$ peaks at 1940 and 2007 eV shown at the right of Figure 2. At the 0.1 eV/ch dispersion necessary to resolve the same Ti-$L_{2,3}$ splitting on the CCD, however, the total energy range is reduced to 204 eV, an 89% decrease from the DED. In order to match the DED's energy range, the CCD must instead be operated at a lower dispersion of 1 eV/channel, sacrificing resolvable fine structure. At the intermediate dispersions, details of the O-K edge and the double peaks of the La-$M_{4,5}$ edge also provide clear benchmarks for comparing PSF between the two detectors, as seen in the middle of Figure 2.

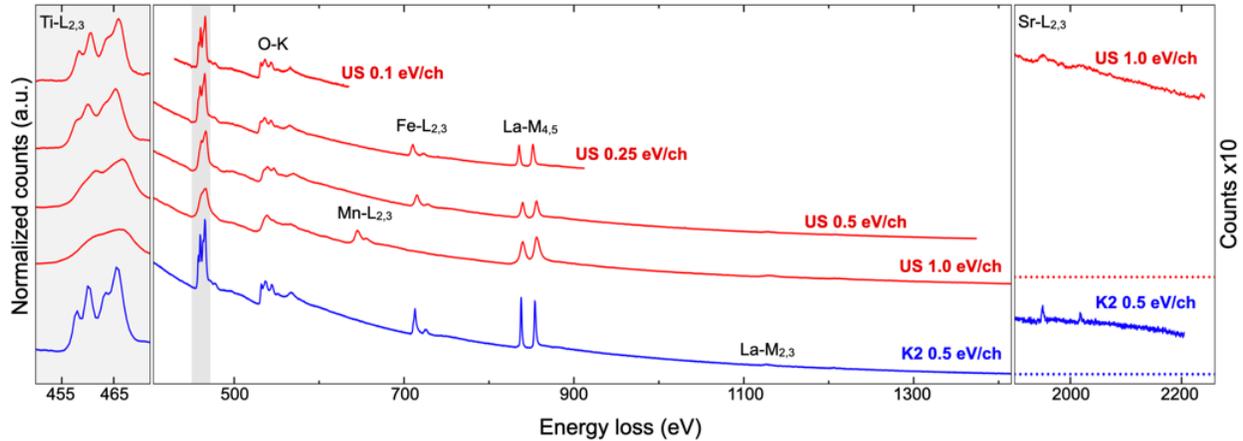

**Figure 2 | Spectral comparison between the two detectors at different dispersions.** Similar energy resolution is observed in the splitting of the Ti-$L_{2,3}$ edge at 0.5 eV/ch on the DED as at 0.1 eV/ch on the CCD (left). In order cover an energy range of ~2000 eV, the CCD must be operated at a low dispersion of 1 eV/channel. Sr-$L_{2,3}$ edges (right) from the same acquisitions are plotted with the zero-count level marked for both spectra by dotted lines; the lower background signal from direct detection increases the jumping ratio of the DED spectra ten times compared to the CCD.

In addition to reducing lateral signal spread, direct detection also eliminates certain sources of background signal in a traditional indirect detector. One major source of background in the CCD is the dark current in the photodiode array (this is distinct from erroneous dark shot noise that affects both detectors, which is corrected by subtracting an averaged dark spectrum after each acquisition). In recorded EEL spectra, the reduction of this background manifests as an improvement to jumping ratio, defined as the quotient of peak signal strength over the pre-edge level. The improvement in jumping ratio can be seen most clearly in relatively low-signal edges, such as the Sr-$L_{2,3}$ edges shown at the right of Figure 2. The zero-count level is marked for each spectrum by the dotted line of corresponding color; jumping ratio improves by a factor of 10 from about 3% on the CCD to over 30% on the DED.

Access to a large energy range and low background are both particularly useful for analyzing complex heterostructures that may contain many edges of interest, possibly separated by hundreds or thousands of eV. Previous work demonstrates the performance of direct EELS detectors at very high energy losses where signal counts are low[30]. Successful mapping of high energy edges, though, is also possible on conventional indirect detectors provided sufficient beam current and dwell times are used so that the signal can overcome the background from the detector readout[31]. In many practical cases, however, the sample constraints or spatial resolution of the experiment limit the total applicable dose, complicating the extraction of low signals from a noisy detector background. Figure 3 shows atomic-resolution elemental mapping of a layered $BiFeO_3$/$LaFeO_3$ (BFO/LFO) superlattice. Taking advantage of the 1855 eV energy range accessible with a 0.5 eV/ch dispersion on the DED, simultaneous maps spanning the minor La-$M_{2,3}$ edge at 1123 eV to the late onset Bi-$M_{4,5}$ edge at 2580 eV can be acquired at atomic resolution using acquisition time as low as 5 ms/px and a modest beam current of <100 pA. The sensitivity and low background of the detector area enables the use of these conditions in order to avoid sample damage and minimize drift effects, both of which could produce artefacts that would inhibit materials analysis.

Short acquisition times over low cross-section edges produce spectra with very few absolute counts, as shown by the summed spectra in Figure 3c. While direct electron detection decreases conversion and readout noise, it also forgoes electron-to-photon conversion gain, so the spectral SNR is instead mostly limited by counting statistics [22,25]. Given $n_i$ counts per channel $i$ per spatial map pixel, Poisson statistics formulate SNR as $\sqrt{n_i}$. The sum of 16 spectra (16 spatial map pixels) from a single Bi atom indicated by the yellow box in the simultaneous ADF of Figure 3a shows a peak of about 10 counts per energy channel at the Bi-$M_{4,5}$ edge. For a single spectrum, the peak signal can thus be inferred $n_i \approx 1$. The noise level described by counting statistics, $\sqrt{n_i}$, is therefore of the same order as the signal! For this reason, comparison between DEDs and higher-gain detectors (such as CCDs) must also take into account other sources of noise and background when comparing detector SNR. Other metrics, such as a signal-to-background or jumping ratio, may in some cases be more appropriate.

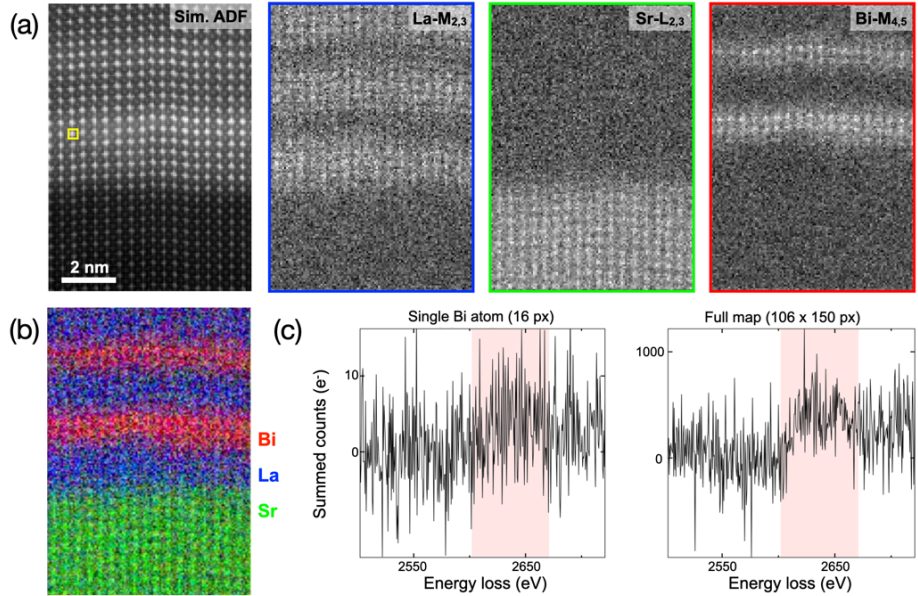

**Figure 3** | Atomic-resolution spectroscopic maps of a BiFeO$_3$ / LaFeO$_3$ superlattice on SrTiO$_3$. (a) Simultaneous ADF image and elemental maps generated from the minor La-M$_{2,3}$ (1123 eV), Sr-L$_{2,3}$ (1940 eV), and Bi-M$_{4,5}$ (2580 eV) edges using 0.5 eV/ch dispersion (1855 eV total range), acquisition time of 5 ms/px, and 100 pA beam current. The full map is 106x150 pixels. (b) False-color RBG image provides detailed information about cation diffusion between layers. (c) Spectra of the Bi-M$_{4,5}$ edge summed over a single Bi column (4×4 px) and the entire SI illustrate the limitation to SNR from counting statistics. An integration window (shaded in red) of 70 eV (140 channels) increases SNR, yielding sufficient contrast for atomic resolution maps.

One important metric for chemical mapping is the SNR of the integrated map. Improving integrated SNR increases the contrast between pixels of a chemical map, enhancing lattice fringe contrast and other metrics of spatial resolution. Particularly for high energy edges, it is possible to improve integrated SNR without increasing the number of counts per channel by integrating over a large energy window, as shown by the red shaded boxes in the spectra of Figure 3c. This increases the total number of counts

$$N = \Sigma n_i$$

that contribute to a single pixel's assigned intensity in a chemical map so that $\sqrt{N}$ is well beyond the level of background noise. For the Bi-$M_{4,5}$ map shown in Figure 3a, even though each channel across a single Bi atom has $n_i \sim 1$ count, an integration window of 70 eV (140 channels) increases the total signal so that $N \gg 1$, producing clearly distinguishable atomic columns.

The benefits of the DED for low-signal mapping suggest its suitability for beam-sensitive samples where the total dose must be limited. Sensitive samples are often studied at lower accelerating voltages in order to reduce the effects of knock-on damage, so the performance of the detector at these conditions is necessarily of interest. Even though the improvements to PSF and effective energy resolution are less significant at 120 kV than at 300 kV (see Figure 1), Figure 4a shows improved mapping contrast on the DED as compared to the CCD for chemical mapping of the Ti-$L_{2,3}$ edge on SrTiO$_3$ for the same map parameters using a primary voltage of 120 kV. As seen in Figure 4b, atomic-resolution mapping of the Sr-$L_{2,3}$ edge at 1940 eV is also achievable on the DED using 30 ms/px dwell time and 50 pA beam current, even with the reduced accelerating voltage. Similar to the Bi-$M_{4,5}$ edge, spectral SNR is again limited by counting statistics with average counts of less than 0.5 e$^-$ per channel per pixel. With a minimum dispersion of 0.25 eV/ch, the benefits of using a large integration (82 eV = 328 ch) window in order to overcome the low-signal Poisson noise are particularly important.

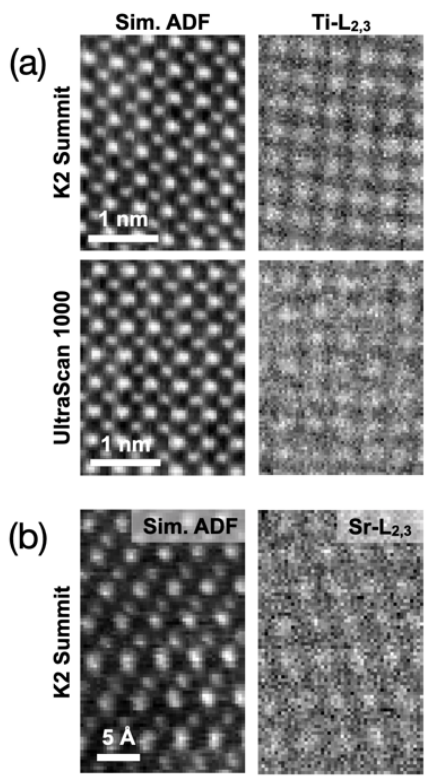

**Figure 4** | Elemental maps of SrTiO$_3$ acquired with low accelerating voltage (120 kV) and low beam current (50 pA). (a) Ti-$L_{2,3}$ maps acquired with 2.5 ms/px dwell times show enhanced atomic contrast in the map acquired by the DED when compared to that acquired on the CCD. (b) Reduced background and noise of direct detection enables atomic resolution mapping with the K2 from high ΔE/E edges with relatively short acquisition times, such as the Sr-$L_{2,3}$ edge mapped with a dwell time of 30 ms/px. Without drift correction, image stability is limited by dwell time.

For edges with large scattering cross sections, the total signal of a spectrum can also be limited simply by very short acquisition time. Figure 5 shows chemical maps of the Ti-$L_{2,3}$, Mn-$L_{2,3}$, and La-$M_{4,5}$ edges extracted from identically parametrized spectrum images of a $La_{0.8}Sr_{0.2}MnO_3$ (LSMO)/$SrTiO_3$ (STO) interface with 2.5 ms/px dwell time acquired on the CCD and the DED. For each element, the DED maps yields lower noise and better contrast, but the improvement is most notable on the Mn-$L_{2,3}$ map (the weakest of the three edges), where fringe contrast in the DED map increases by 40% over the CCD. In similar experiments with dwell times of 5 and 10 ms/px, the difference between integration maps from the two detectors is less noticeable, although the corresponding spectra show similar differences in spectral effective energy resolution as discussed above (see Figure 2). The most notable difference between the data sets, though, is not contrast or noise but total SI time: although the two maps are the same size and used the same per-pixel dwell time, the CCD map took 22% longer to acquire than the DED due to a 4-fold increase in readout time per frame. On average, the DED has a dead time of 0.15 ms/px compared to 0.62 ms/px on the CCD, an important metric for variable temperature experiments typically limited by stage stability and drift[32,33,34].

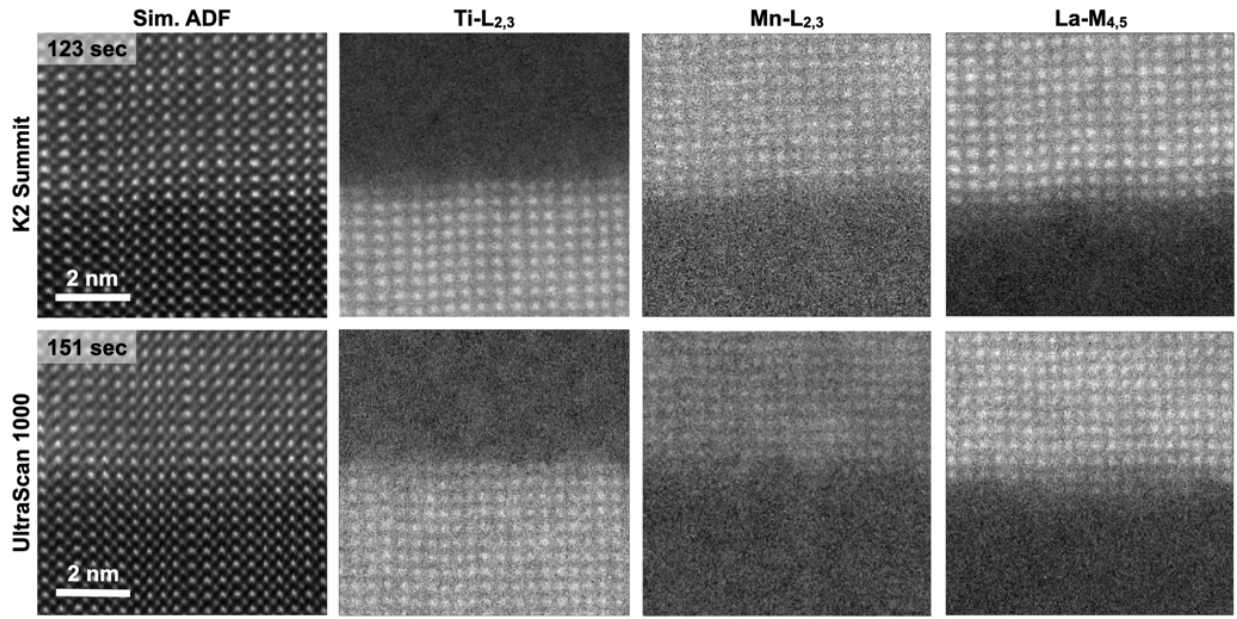

**Figure 5** | Elemental maps of the Ti-$L_{2,3}$, Mn-$L_{2,3}$, and La-$M_{4,5}$ edges extracted from identically acquired spectrum images of a $(La_{0.8}Sr_{0.2})MnO_3$/$SrTiO_3$ interface with 2.5 ms/px dwell time. Fringe contrast of the Mn-$L_{2,3}$ map increases by 40% on the DED compared to the CCD. Maps are the same size with the same per pixel dwell time; the CCD map took 22% longer to record than the DED map due to a four-fold increase in readout time per frame.

**DISCUSSION**

It is worth noting that despite the advantages of direct detection for low-signal experimental conditions, there remain many applications in which traditional detection is equivalent or even preferable. Currently available direct detectors are limited by pixel saturation: while it is possible to reduce large signals with lower duty cycle, this will also attenuate small signals by the same factor. Other compatibilities such as Dual EELS will also need to be integrated for DED. As such, a CCD or other camera with larger dynamic range may be a better choice for low-loss or other experiments with signals of widely varying relative strength.

Still, compared to an indirect detector, the DED offers select key advantages for low-signal EELS mapping: improved spectrum quality with low background and narrow PSF, high energy resolution retained over wide simultaneous ranges, and fast readout with low per-frame dead time. The ability to map from spectra with low counts is useful for a variety of applications, including dose-sensitive specimens, low cross section edges, and short acquisition times. Low-signal performance combined with access to a simultaneous energy range of nearly 2000 eV affords new flexibility for experiments to simultaneously probe multiple edges in the same structure. Together with the low dead time, all of these factors enable high quality chemical maps to be acquired in significantly less time than on a traditional detector.

Figure 6 shows atomic-resolution elemental mapping of the LSMO/STO interface at cryogenic temperatures near 96 K enabled by the rapid acquisition spectrum imaging of the DED. Even after settling for multiple hours, the Gatan 636 side-entry liquid nitrogen cryo-holder experiences significant reduction in stability compared to room temperature due to thermal drift and bubbling in the nitrogen[34]. Previous experiments have successfully used EELS to track electronic and chemical changes in bulk samples and across interfaces at temperatures as low as 10 K, but stage instabilities have until now prevented two-dimensional atomic resolution mapping[32,35,36]. To minimize skewing and tearing of the resulting maps, it is therefore imperative to acquire spectrum images in as little total time as possible. Because of its fast frame rate and low dead time, a 48×268 px map with 2.5 ms/px dwell time takes only 35 sec to acquire on the DED while preserving signal quality sufficient to distinguish single layers of Ti and La diffusion at the interface.

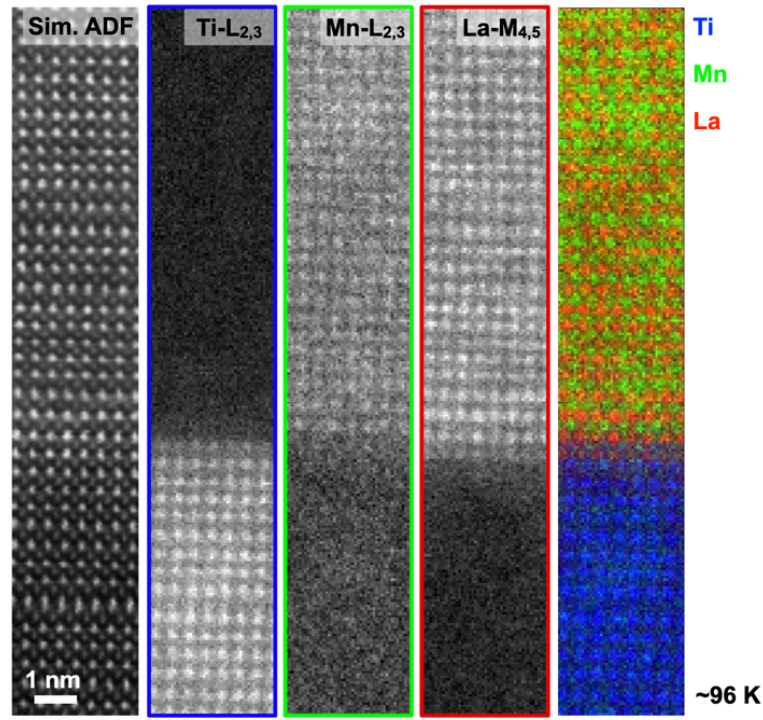

**Figure 6** | Atomic-resolution elemental mapping of a $(La_{0.8}Sr_{0.2})MnO_3/SrTiO_3$ interface at cryogenic temperature (~96 K), enabled by rapid acquisition spectrum imaging with the DED. A 2.5 ms/px dwell time was used to minimize distortions due to drift, though some artefacts are visible along the slow (vertical) scan direction. The entire 48x268 px map took a total of 35 sec to acquire.

High resolution cryogenic EELS mapping opens the door to new experiments across many fields. In materials physics, low temperature measurements can probe exotic phases such as charge ordering and metal-to-insulator transitions[32,35,37]. Other systems must be imaged at cryogenic temperatures in order to conduct an experiment at all: for reactive grain boundaries in 2D materials or nanoparticles that are not robust to standard cleaning methods, cryogenic imaging can reduce or prevent carbon contamination in the column. On larger lengths scales, the increased sensitivity of direct detection will help expend EELS to dose limited cryo-immobilized samples, including hydrated biological structures[14], solid-liquid interfaces[13], and organic-mineral interfaces.

**CONCLUSIONS**

Originally developed for high-resolution, low-dose TEM imaging, the Gatan K2 Summit direct electron detector offers many advantages for a number of EELS applications. The process of direct detection yields narrower PSF, lower background, and reduced noise contributions as compared to a traditional indirect detector. Together, these improvements enhance the sensitivity of EELS experiments, providing greater flexibility to encompass large energy ranges, low cross-section edges, and dose limitations. Furthermore, the fast frame rate and very low per-pixel readout dead time of the K2 DED enable acquisition of high quality spectrum images in much shorter times than the UltraScan CCD. Taking advantage of direct detection, we used the DED to acquire atomic-resolution chemical maps at cryogenic temperatures without any drift correction or other post-processing. The demonstration of high resolution EELS mapping at cryogenic and other low-signal conditions by direct detection is a promising development for the expansion of the technique to new fields across materials science.

## ACKNOWLEDGEMENTS

Oxide samples generously provided by Julia A. Mundy, Ramamoorthy Ramesh, Yasuyuki Hikita, and Harold Y. Hwang.

This work was primarily supported by the National Sciences Foundation, through the PARADIM Materials Innovation Platform (DMR-1539918). We acknowledge additional support by the Department of Defense Air Force Office of Scientific Research (No. FA 9550-16-1-0305) and the Packard Foundation. This work made use of the Cornell Center for Materials Research (CCMR) Shared Facilities, which are supported through the NSF MRSEC Program (No. DMR-1719875). The FEI Titan Themis 300 was acquired through No. NSF-MRI-1429155, with additional support from Cornell University, the Weill Institute, and the Kavli Institute at Cornell.

## COMPETING INTERESTS

The authors declare that they have no competing financial interests.

## MATERIALS & CORRESPONDENCE

Correspondence and requests for materials should be addressed to L.F.K. (lena.f.kourkoutis@cornell.edu).